\documentclass[12pt]{article}
\usepackage{amssymb}

\usepackage{graphicx}
\usepackage{amsmath}

\input{tcilatex}

\begin{document}

\title{On the Debye effect in a semiconductor and an electrolyte}
\author{S. Piekarski \\
IPPT PAN}
\maketitle

\begin{abstract}
It is shown that the standard expression for the diffusive flux in a
semiconductor can be transformed to the flux from the Einstein -
Smoluchowski equation after making use of the ''Einstein identities''.
Later, it is possible to use a transformation introduced in: S.Piekarski,
''On the modified Fick law and its potential applications''
(J.Tech.Phys.,44,2,125-131,2003). In this formalism, it is possible to
formulate simple models of a semiconductor and an electrolite and to obtain
the alternative description of a Debye effect (by the Debye effect we mean
here the screening of the point particle potential). Final results are
compared with the standard analysis.
\end{abstract}

\section{Introduction\protect\bigskip}

According to Peierls ($\left[ 1\right] $, p.204, Eq.$\left( 10.21\right) $),
the electric current $J$ in a semiconductor is given by 
\begin{equation}
J=-\sigma \frac{\partial \phi }{\partial x}-D\frac{\partial n}{\partial x},
\end{equation}
where $\sigma $ is the conductivity, $D$ is the diffusion coefficient, $\phi 
$ is the electric potential and $n$ is the concentration of the electrons.
The conductivity $\sigma $ and the diffusion coefficient $D$ are related by
the ''Einstein identity'' 
\begin{equation}
D=\frac{kT}{e}\frac{\sigma }{n}=kTu,  \label{'}
\end{equation}
where $u=\sigma /ne$ is independent of the electron density, and is called
the ''mobility'' (compare also $\left[ 5\right] $, p.199).

From the ''Einstein identity'' $\left( 2\right) $ one can determine the
conductivity $\sigma $ as a function of the temperature and the diffusion
coefficient 
\begin{equation}
\sigma =\frac{eDn}{kT}.
\end{equation}
After inserting $\left( 3\right) $ into $\left( 1\right) $ one arrives at 
\begin{equation}
J=-\sigma \frac{\partial \phi }{\partial x}-D\frac{\partial n}{\partial x}=-%
\frac{eDn}{kT}\frac{\partial \phi }{\partial x}-D\frac{\partial n}{\partial x%
}=-D\left[ \frac{\partial n}{\partial x}+\frac{en}{kT}\frac{\partial \phi }{%
\partial x}\right] .
\end{equation}
It can be seen that $\left( 4\right) $ is a flux from the Einstein -
Smoluchowski equation (up to minor differencies in the conventions for signs
and notations) $\left[ 2,3,8,9\right] $. For the constant temperature field,
this flux can be written in the following form $\left[ 2,3\right] $%
\begin{equation}
J=-D\exp \left[ -\frac{e\phi }{kT}\right] \frac{\partial }{\partial x}%
\left\{ n\exp \left[ \frac{e\phi }{kT}\right] \right\} .
\end{equation}
Let us show this explicitly 
\begin{equation*}
-D\exp \left[ -\frac{e\phi }{kT}\right] \frac{\partial }{\partial x}\left\{
n\exp \left[ \frac{e\phi }{kT}\right] \right\} =
\end{equation*}
\begin{equation*}
-D\exp \left[ -\frac{e\phi }{kT}\right] \exp \left[ \frac{e\phi }{kT}\right] 
\frac{\partial }{\partial x}n-D\exp \left[ -\frac{e\phi }{kT}\right] n\frac{%
\partial }{\partial x}\exp \left[ \frac{e\phi }{kT}\right] =
\end{equation*}
\begin{equation*}
-D\frac{\partial }{\partial x}n-D\exp \left[ -\frac{e\phi }{kT}\right] \exp %
\left[ \frac{e\phi }{kT}\right] n\frac{\partial }{\partial x}\left[ \frac{%
e\phi }{kT}\right] =
\end{equation*}
\begin{equation}
-D\frac{\partial }{\partial x}n-\frac{Dne}{kT}\frac{\partial }{\partial x}%
\phi =-D\left[ \frac{\partial }{\partial x}n+\frac{ne}{kT}\frac{\partial }{%
\partial x}\phi \right] =J.
\end{equation}
It should be stressed again that the temperature $T$ in $\left( 5\right) $
and $\left( 6\right) $ is constant.

On p.204 of $\left[ 1\right] $ Peierls writes that the concentration of
electrons is given by the ''Boltzmann formula'' if the system is in
equilibrium in the spatially varying electric potential $\phi =\phi \left(
x\right) $. This property can be shown immediately after putting the flux $%
\left( 5\right) $ equal to zero: 
\begin{equation}
J=-D\exp \left[ -\frac{e\phi }{kT}\right] \frac{\partial }{\partial x}%
\left\{ n\exp \left[ \frac{e\phi }{kT}\right] \right\} =0.
\end{equation}
Since the diffusion coefficient is assumed to be different from zero, the
above condition implies that 
\begin{equation}
\frac{\partial }{\partial x}\left\{ n\left( x\right) \exp \left[ \frac{e\phi
\left( x\right) }{kT}\right] \right\} =0
\end{equation}
and therefore 
\begin{equation}
n\left( x\right) =C\exp \left[ -\frac{e\phi \left( x\right) }{kT}\right] ,%
\text{ \ \ \ \ \ \ \ \ }C>0.
\end{equation}

It is worth to discuss also the explicit form of the Einstein - Smoluchowski
equation for the electrically neutral matter; in $\left[ 2,3\right] $, the
following convention has been used$:$%
\begin{equation}
\frac{\partial }{\partial t}p\left( x,t\right) =\frac{\partial }{\partial x}%
D^{\prime }\left( T,x\right) \left\{ \frac{\partial }{\partial x}p\left(
x,t\right) +\frac{p\left( x,t\right) }{kT}\frac{\partial }{\partial x}%
U\left( x\right) \right\} ,
\end{equation}
where $U\left( x\right) $ is a potential for the external force acting on
the diffusing molecule, $p$ is the concentration of the diffusing molecules,
and $D^{\prime }\left( T,x\right) $ is the diffusion coefficient. The
equation $\left( 10\right) $ can be generalized to the equation for
diffusion of the electrically charged matter after the replacement 
\begin{equation}
U\left( x\right) \rightarrow U\left( x\right) -q\phi \left( x,t\right)
\end{equation}
where $q$ is the electric charge of a diffusing molecule and $\phi \left(
x,t\right) $ denotes the electric potential $\left[ 4\right] $. The flux of
diffusing molecules, corresponding to the equation 
\begin{equation}
\frac{\partial }{\partial t}p\left( x,t\right) =\frac{\partial }{\partial x}%
D^{\prime }\left( T,x\right) \left\{ \frac{\partial }{\partial x}p\left(
x,t\right) +\frac{p\left( x,t\right) }{kT}\frac{\partial }{\partial x}\left[
U\left( x\right) -q\phi \left( x,t\right) \right] \right\} ,
\end{equation}
is given by 
\begin{equation}
J^{\prime }=D^{\prime }\left( T,x\right) \left\{ \frac{\partial }{\partial x}%
p\left( x,t\right) +\frac{p\left( x,t\right) }{kT}\frac{\partial }{\partial x%
}\left[ U\left( x\right) -q\phi \left( x,t\right) \right] \right\} .
\end{equation}
In turn, the electric flux, corresponding to the flux of molecules $\left(
13\right) ,$ is equal to the product of the flux of molecules $\left(
13\right) $ and the electric charge of a single molecule $q:$%
\begin{equation}
J_{e}^{\prime }=qD^{\prime }\left( T,x\right) \left\{ \frac{\partial }{%
\partial x}p\left( x,t\right) +\frac{p\left( x,t\right) }{kT}\frac{\partial 
}{\partial x}\left[ U\left( x\right) -q\phi \left( x,t\right) \right]
\right\} .
\end{equation}
In particular, if the non - electric potential $U\left( x\right) $ is
constant 
\begin{equation}
U\left( x\right) =U_{0}=const.
\end{equation}
the electric flux $\left( 14\right) $ takes the form 
\begin{equation*}
J_{e}^{\prime }=qD^{\prime }\left( T,x\right) \left\{ \frac{\partial }{%
\partial x}p\left( x,t\right) +\frac{p\left( x,t\right) }{kT}\frac{\partial 
}{\partial x}\left[ U_{0}-q\phi \left( x,t\right) \right] \right\} =
\end{equation*}
\begin{equation*}
qD^{\prime }\left( T,x\right) \left\{ \frac{\partial }{\partial x}p\left(
x,t\right) +\frac{p\left( x,t\right) }{kT}\frac{\partial }{\partial x}\left[
-q\phi \left( x,t\right) \right] \right\} =
\end{equation*}
\begin{equation}
qD^{\prime }\left( T,x\right) \frac{\partial }{\partial x}p\left( x,t\right)
-q^{2}D^{\prime }\left( T,x\right) \frac{p\left( x,t\right) }{kT}\frac{%
\partial }{\partial x}\phi \left( x,t\right) .
\end{equation}
The electric flux $\left( 16\right) $ is identical to the electric flux from
the Peierls formula under the conditions 
\begin{equation}
-\frac{eDn}{kT}\frac{\partial }{\partial x}\phi \left( x,t\right) =-\sigma 
\frac{\partial }{\partial x}\phi \left( x,t\right) =-q^{2}D^{\prime }\left(
T,x\right) \frac{p\left( x,t\right) }{kT}\frac{\partial }{\partial x}\phi
\left( x,t\right)
\end{equation}
and 
\begin{equation}
-D\frac{\partial }{\partial x}n=qD^{\prime }\left( T,x\right) \frac{\partial 
}{\partial x}p\left( x,t\right) ,
\end{equation}
(in $\left( 17\right) $, the ''Einstein identity'' $\left( 3\right) $ has
been taken into account). It implies that 
\begin{equation*}
n=p
\end{equation*}
and 
\begin{equation*}
e=q.
\end{equation*}
From the relation 
\begin{equation}
-D=qD^{\prime }\left( T,x\right)
\end{equation}
one can see that there is a difference in notational conventions for the
diffusion coefficients. In the Peierls' version, the charge of the carrier
is inserted into the definition of the diffusion coefficient while in $%
\left( 14\right) $ they are explicitly separated. It seems useful to compare
both fluxes with the balance of electric charge, derived from Maxwell
equations. In general, the electric flux is a part of Maxwell equations 
\begin{equation}
rotH=j+\frac{\partial }{\partial t}D_{e}
\end{equation}
\begin{equation}
rotE=-\frac{\partial }{\partial t}B
\end{equation}
\begin{equation}
divB=0
\end{equation}
\begin{equation}
divD_{e}=\rho _{s}.
\end{equation}
After taking a divergence of $\left( 20\right) $ and the time derivative of $%
\left( 23\right) $ one arrives at the identities 
\begin{equation}
0=divj+\frac{\partial }{\partial t}divD_{e}
\end{equation}
and 
\begin{equation}
\frac{\partial }{\partial t}divD_{e}=\frac{\partial }{\partial t}\rho _{s}.
\end{equation}
Comparing of $\left( 24\right) $ and $\left( 25\right) $ gives 
\begin{equation}
\frac{\partial }{\partial t}\rho _{s}+divj=0.
\end{equation}
From the form of $\left( 26\right) $ one can see that the sign convention
applied in $\left( 14\right) $ is opposite to that used in $\left( 20\right) 
$ (compare $\left[ 2,3\right] $). In the rest of this paper we shall use the
notational convention used in $\left( 14\right) $, but without the prime in
the notation for the diffusion constant.

In order to discuss the Debye effect in a semiconductor, let us assume that
the electric charge $\rho _{s}$ is a sum of the electric charge of the
negative carriers $en\left( x,t\right) $ and the electric charge density of
the static ''background'' $-e\rho _{b}$. The processes of ionization and
recombination are neglected. Since we are interested in the Debye effect, we
assume that, besides the constant background charge, there is a point charge 
$\eta e\delta \left( x\right) $ at $x=0$ Therefore, the final expression for 
$\rho _{s}$ is 
\begin{equation}
\rho _{s}=en\left( x,t\right) -e\rho _{b}+\eta e\delta \left( x\right)
\end{equation}
while the explicit form of the diffusive flux in Maxwell equations is 
\begin{equation}
j=-eD\left( x,T\right) \exp \left[ -\frac{n\phi }{kT}\right] \frac{\partial 
}{\partial x}\left\{ n\exp \left[ \frac{n\phi }{kT}\right] \right\} .
\end{equation}
Under these assumptions, Maxwell equations become 
\begin{equation}
rotH=-eD\exp \left[ -\frac{e\phi }{kT}\right] \frac{\partial }{\partial x}%
\left\{ n\exp \left[ \frac{e\phi }{kT}\right] \right\} +\frac{\partial }{%
\partial t}D_{e},
\end{equation}

\begin{equation}
rotE=-\frac{\partial }{\partial t}B,
\end{equation}

\begin{equation}
divB=0,
\end{equation}

\begin{equation}
divD_{e}=en\left( x,t\right) -e\rho _{b}+\eta e\delta \left( x\right) ,
\end{equation}

and the balance of the electric charge becomes 
\begin{equation*}
\frac{\partial }{\partial t}\left[ en\left( x,t\right) -e\rho _{b}+\eta
e\delta \left( x\right) \right] +div\left\{ -eD\exp \left[ -\frac{e\phi }{kT}%
\right] \frac{\partial }{\partial x}\left\{ n\exp \left[ \frac{e\phi }{kT}%
\right] \right\} \right\} =
\end{equation*}

\begin{equation}
\frac{\partial }{\partial t}en\left( x,t\right) +div\left\{ -eD\exp \left[ -%
\frac{e\phi }{kT}\right] \frac{\partial }{\partial x}\left\{ n\exp \left[ 
\frac{e\phi }{kT}\right] \right\} \right\} =0
\end{equation}
For simplicity, one can assume that

\begin{equation}
D_{e}=\varepsilon _{0}\varepsilon _{r}E,
\end{equation}

\begin{equation}
B=\mu _{0}\mu _{r}H,
\end{equation}
and 
\begin{equation}
B=0.
\end{equation}
In turn, $\left( 30\right) $ and $\left( 36\right) $ imply that 
\begin{equation}
rotE=0.
\end{equation}
If 
\begin{equation}
E=-\nabla \phi \left( x,t\right)
\end{equation}
then the remaining equations take the form 
\begin{equation}
0=-eD\exp \left[ -\frac{e\phi }{kT}\right] \frac{\partial }{\partial x}%
\left\{ n\exp \left[ \frac{e\phi }{kT}\right] \right\} +\frac{\partial }{%
\partial t}\varepsilon _{0}\varepsilon _{r}\left[ -\nabla \phi \right] ,
\end{equation}
and 
\begin{equation}
-\varepsilon _{0}\varepsilon _{r}\nabla ^{2}\phi =en\left( x,t\right) -e\rho
_{b}+\eta e\delta \left( x\right) .
\end{equation}
In order to describe the process of dynamical screening one has to impose
the corresponding initial and boundary conditions. Let us assume that the
support of the diffusion coefficient (or, strictly speaking, the tensor of
the diffusion coefficients) is restricted to the interior of the sphere with
a radius $R$. The important class of solutions of $\left( 39\right) $ and $%
\left( 40\right) $ consists of stationary solutions for which the time
derivative is put equal to zero 
\begin{equation}
0=-eD\exp \left[ -\frac{e\phi }{kT}\right] \frac{\partial }{\partial x}%
\left\{ n\exp \left[ \frac{e\phi }{kT}\right] \right\} ,\text{ \ \ \ \ \ \ \ 
}\frac{\partial \phi }{\partial t}=0,
\end{equation}
\begin{equation}
-\varepsilon _{0}\varepsilon _{r}\nabla ^{2}\phi =en\left( x\right) -e\rho
_{b}+\eta e\delta \left( x\right) .
\end{equation}

The condition $\left( 41\right) _{1}$ can be solved to give 
\begin{equation}
n\left( x\right) =C\exp \left[ -\frac{e\phi \left( x\right) }{kT}\right] ,%
\text{ \ \ \ \ \ \ \ },C>0
\end{equation}
(compare $\left( 7\right) -\left( 9\right) $). In principle, the value of
the positive paremeter $C$ can be determined from the initial conditions.

Now, $\left( 43\right) $ can be inserted into $\left( 42\right) $: 
\begin{equation}
-\varepsilon _{0}\varepsilon _{r}\nabla ^{2}\phi \left( x\right) =eC\exp 
\left[ -\frac{e\phi \left( x\right) }{kT}\right] -e\rho _{b}+\eta e\delta
\left( x\right) .
\end{equation}
This equation depends explicitly on the three parameters $C,\rho _{b}$ and $%
\eta $. The fourth parameter is the radius $R$ of the sphere in which the
transport processes take place. In the more detailed analysis one should use
the dimensionless variables, taking into accunt the spatial scales of the
system (for example, the radius $R$ ), in the way similar to that used in
the Chapman - Enskog expansion, but in our preliminary discussion we simply
assume that 
\begin{equation}
\exp \left[ -\frac{e\phi \left( x\right) }{kT}\right] \thickapprox 1-\frac{%
e\phi \left( x\right) }{kT}.
\end{equation}
After inserting $\left( 45\right) $ into $\left( 44\right) $ one arrives at 
\begin{equation}
-\varepsilon _{0}\varepsilon _{r}\nabla ^{2}\phi \left( x\right) =eC\left[ 1-%
\frac{e\phi \left( x\right) }{kT}\right] -e\rho _{b}+\eta e\delta \left(
x\right) .
\end{equation}
It is well - known that the solution of the modified Poisson equation 
\begin{equation}
\left( \nabla ^{2}-\mu ^{2}\right) \phi =-4\pi e\delta \left( r\right)
\end{equation}
is of the form 
\begin{equation}
\phi \left( r\right) =\frac{e}{r}\exp \left( -\mu r\right) .
\end{equation}
Some solutions of $\left( 46\right) $ can be determined from the solution $%
\left( 48\right) $ of the modified Poisson equation $\left( 47\right) $
after the corresponding change of variables. Let us describe it in detail.

Eq.$\left( 46\right) $ is equivalent to 
\begin{equation}
\left( \nabla ^{2}-\frac{e^{2}C}{kT\varepsilon _{0}\varepsilon _{r}}\right)
\phi \left( x\right) =-\frac{e}{\varepsilon _{0}\varepsilon _{r}}\left[
C-\rho _{b}\right] -\frac{\eta e}{\varepsilon _{0}\varepsilon _{r}}\delta
\left( x\right) .
\end{equation}
Therefore, one can introduce the notation 
\begin{equation}
\mu ^{2}=\frac{e^{2}C}{kT\varepsilon _{0}\varepsilon _{r}}.
\end{equation}
Now, we would like to change the variables in 
\begin{equation}
\left( \nabla ^{2}-\mu ^{2}\right) \phi \left( x\right) =-\frac{e}{%
\varepsilon _{0}\varepsilon _{r}}\left[ C-\rho _{b}\right] -\frac{\eta e}{%
\varepsilon _{0}\varepsilon _{r}}\delta \left( x\right) ,
\end{equation}
transforming it into the modified Poisson equation. To this aim, we insert 
\begin{equation}
\phi \left( x\right) =\phi ^{\prime }\left( x\right) +\alpha
\end{equation}
into the l.h.s. of $\left( 51\right) $: 
\begin{equation}
\left( \nabla ^{2}-\mu ^{2}\right) \left[ \phi ^{\prime }\left( x\right)
+\alpha \right] =\left( \nabla ^{2}-\mu ^{2}\right) \phi ^{\prime }\left(
x\right) -\mu ^{2}\alpha =-\frac{e}{\varepsilon _{0}\varepsilon _{r}}\left[
C-\rho _{b}\right] -\frac{\eta e}{\varepsilon _{0}\varepsilon _{r}}\delta
\left( x\right) .
\end{equation}
Eq.$\left( 53\right) $ is equivalent to 
\begin{equation}
\left( \nabla ^{2}-\mu ^{2}\right) \phi ^{\prime }\left( x\right) =\mu
^{2}\alpha -\frac{e}{\varepsilon _{0}\varepsilon _{r}}\left[ C-\rho _{b}%
\right] -\frac{\eta e}{\varepsilon _{0}\varepsilon _{r}}\delta \left(
x\right) ,
\end{equation}
and the value of $\alpha $ can be determined from the condition that the sum
of first two terms on the r.h.s. of $\left( 54\right) $ should vanish. It
means that 
\begin{equation}
\mu ^{2}\alpha -\frac{e}{\varepsilon _{0}\varepsilon _{r}}\left[ C-\rho _{b}%
\right] =0
\end{equation}
what after insertion of $\left( 50\right) $ into $\left( 55\right) $ gives 
\begin{equation}
\alpha =\frac{kT}{eC}\left[ C-\rho _{b}\right] .
\end{equation}
Therefore, our equation $\left( 51\right) $ takes the form 
\begin{equation}
\left[ \nabla ^{2}-\mu ^{2}\right] \phi ^{\prime }\left( x\right) =-\frac{%
\eta e}{\varepsilon _{0}\varepsilon _{r}}\delta \left( x\right) .
\end{equation}
In order to transform $\left( 57\right) $ into the ''modified Poisson
equation'' one can make the following transformation 
\begin{equation}
\left[ \nabla ^{2}-\mu ^{2}\right] \phi ^{\prime }\left( x\right) =-\frac{%
\eta e}{\varepsilon _{0}\varepsilon _{r}}\delta \left( x\right) =-4\pi \frac{%
\eta e}{4\pi \varepsilon _{0}\varepsilon _{r}}\delta \left( x\right) .
\end{equation}
After comparing $\left( 58\right) $ and $\left( 47\right) $ one sees that 
\begin{equation}
\phi ^{\prime }\left( x\right) =\frac{\eta e}{4\pi \varepsilon
_{0}\varepsilon _{r}x}\exp \left[ -\left( \frac{e^{2}C}{kT\varepsilon
_{0}\varepsilon _{r}}\right) ^{\frac{1}{2}}x\right]
\end{equation}
and therefore 
\begin{equation}
\phi \left( x\right) =\phi ^{\prime }\left( x\right) +\alpha =\frac{\eta e}{%
4\pi \varepsilon _{0}\varepsilon _{r}x}\exp \left[ -\left( \frac{e^{2}C}{%
kT\varepsilon _{0}\varepsilon _{r}}\right) ^{\frac{1}{2}}x\right] +\frac{kT}{%
eC}\left[ C-\rho _{b}\right] .
\end{equation}
A similar discussion can be made in the case of an electrolyte.

Again we start with the Maxwell equations but in thie case of an electrolyte
we introduce two different kind of carriers, one charged with a negative
charge and the second one charged with a positive charge. The density of the
negative carriers is denoted $n\left( x,t\right) $ while the density of the
positive carriers is denoted $p\left( x,t\right) $. We also introduce two
different fluxes 
\begin{equation}
j_{n}=-eD_{n}\exp \left[ -\frac{e\phi }{kT}\right] \frac{\partial }{\partial
x}\left\{ n\exp \left[ \frac{e\phi }{kT}\right] \right\}
\end{equation}
and 
\begin{equation}
j_{p}=eD_{p}\exp \left[ \frac{e\phi }{kT}\right] \frac{\partial }{\partial x}%
\left\{ p\exp \left[ -\frac{e\phi }{kT}\right] \right\} .
\end{equation}

Also in this case we introduce the point charge $\eta e\delta \left(
x\right) $ fixed at the point $x=0$. We also assume that the static
background charge is absent in this case and the processes of ionization and
recombination are absent. Therefore, Maxwell equations are in this case 
\begin{equation*}
rotH=-eD_{n}\exp \left[ -\frac{e\phi }{kT}\right] \frac{\partial }{\partial x%
}\left\{ n\exp \left[ \frac{e\phi }{kT}\right] \right\} +
\end{equation*}
\begin{equation}
eD_{p}\exp \left[ \frac{e\phi }{kT}\right] \frac{\partial }{\partial x}%
\left\{ p\exp \left[ -\frac{e\phi }{kT}\right] \right\} +\frac{\partial }{%
\partial t}D_{e}
\end{equation}
\begin{equation}
rotE=-\frac{\partial }{\partial t}B
\end{equation}
\begin{equation}
divB=0
\end{equation}
\begin{equation}
divD_{e}=en\left( x,t\right) -ep\left( x,t\right) +\eta e\delta \left(
x\right) .
\end{equation}
Our futher discussion is analogous to the previously discussed case of a
semiconductor. We still assume $\left( 34\right) $, $\left( 35\right) $, $%
\left( 36\right) $ and obtain $\left( 37\right) $ as a consequence. Then, we
assume $\left( 38\right) $ and obtain, as the counterparts of $\left(
39\right) $ and $\left( 40\right) $, equations 
\begin{equation*}
0=-eD_{n}\exp \left[ -\frac{e\phi }{kT}\right] \frac{\partial }{\partial x}%
\left\{ n\exp \left[ \frac{e\phi }{kT}\right] \right\} +
\end{equation*}
\begin{equation}
eD_{p}\exp \left[ \frac{e\phi }{kT}\right] \frac{\partial }{\partial x}%
\left\{ p\exp \left[ -\frac{e\phi }{kT}\right] \right\} +\frac{\partial }{%
\partial t}\left[ -\varepsilon _{0}\varepsilon _{r}\nabla \phi \right]
\end{equation}
and 
\begin{equation}
-\varepsilon _{0}\varepsilon _{r}\nabla ^{2}\phi \left( x,t\right) =en\left(
x,t\right) -ep\left( x,t\right) +\eta e\delta \left( x\right) .
\end{equation}
We discuss the particular solutions of $\left( 67\right) $ and $\left(
68\right) $, with the time derivative equal to zero and both fluxes equal to
zero 
\begin{equation}
-eD_{n}\exp \left[ -\frac{e\phi }{kT}\right] \frac{\partial }{\partial x}%
\left\{ n\exp \left[ \frac{e\phi }{kT}\right] \right\} =0
\end{equation}
\begin{equation}
eD_{p}\exp \left[ \frac{e\phi }{kT}\right] \frac{\partial }{\partial x}%
\left\{ p\exp \left[ -\frac{e\phi }{kT}\right] \right\} =0.
\end{equation}

Similarly as before, we solve $\left( 69\right) $ and $\left( 70\right) $
and insert the results into $\left( 68\right) $, arriving at 
\begin{equation}
-\varepsilon _{0}\varepsilon _{r}\nabla ^{2}\phi \left( x\right) =eC_{n}\exp 
\left[ -\frac{e\phi }{kT}\right] -eC_{p}\exp \left[ \frac{e\phi }{kT}\right]
+\eta e\delta \left( x\right)
\end{equation}
this time. Both exponents from $\left( 71\right) $ are linearized according
to the rules 
\begin{equation}
\exp \left[ -\frac{e\phi }{kT}\right] \thickapprox 1-\frac{e\phi }{kT}
\end{equation}
\begin{equation}
\exp \left[ \frac{e\phi }{kT}\right] \thickapprox 1+\frac{e\phi }{kT}
\end{equation}
and therefore one arrives at 
\begin{equation}
-\varepsilon _{0}\varepsilon _{r}\nabla ^{2}\phi \left( x\right) =eC_{n} 
\left[ 1-\frac{e\phi }{kT}\right] -eC_{p}\left[ 1+\frac{e\phi }{kT}\right]
+\eta e\delta \left( x\right) .
\end{equation}

This equation is equivalent to

\begin{equation}
\left[ \nabla ^{2}-\left( \frac{e^{2}}{\varepsilon _{0}\varepsilon _{r}kT}%
\left[ C_{n}+C_{p}\right] \right) \right] \phi \left( x\right) =-\frac{e}{%
\varepsilon _{0}\varepsilon _{r}}\left[ C_{n}-C_{p}\right] -\frac{\eta e}{%
\varepsilon _{0}\varepsilon _{r}}\delta \left( x\right) .
\end{equation}
Similarly as in the case of a semiconductor $\left( 50\right) $, one defines
the parameter $\mu ^{2}$ which is now 
\begin{equation}
\mu ^{2}=\frac{e^{2}}{\varepsilon _{0}\varepsilon _{r}kT}\left[ C_{n}+C_{p}%
\right]
\end{equation}
and after inserting $\left( 76\right) $ into $\left( 75\right) $ one arrives
at 
\begin{equation}
\left[ \nabla ^{2}-\mu ^{2}\right] \phi \left( x\right) =-\frac{e}{%
\varepsilon _{0}\varepsilon _{r}}\left[ C_{n}-C_{p}\right] -\frac{\eta e}{%
\varepsilon _{0}\varepsilon _{r}}\delta \left( x\right) .
\end{equation}
After inserting $\left( 52\right) $ into the l.h.s. of $\left( 77\right) $
one arrives at 
\begin{equation}
\left[ \nabla ^{2}-\mu ^{2}\right] \phi ^{\prime }\left( x\right) -\mu
^{2}\alpha =-\frac{e}{\varepsilon _{0}\varepsilon _{r}}\left[ C_{n}-C_{p}%
\right] -\frac{\eta e}{\varepsilon _{0}\varepsilon _{r}}\delta \left(
x\right) .
\end{equation}
The above furmula is an analogy to $\left( 53\right) $. Now, $\left(
78\right) $ is equivalent to 
\begin{equation}
\left[ \nabla ^{2}-\mu ^{2}\right] \phi ^{\prime }\left( x\right) =\mu
^{2}\alpha -\frac{e}{\varepsilon _{0}\varepsilon _{r}}\left[ C_{n}-C_{p}%
\right] -\frac{\eta e}{\varepsilon _{0}\varepsilon _{r}}\delta \left(
x\right)
\end{equation}
and the value of $\alpha $ is determined here from the condition that the
sum of first two terms on the r.h.s. of $\left( 79\right) $ should vanish.
Therefore 
\begin{equation}
\mu ^{2}\alpha =\frac{e}{\varepsilon _{0}\varepsilon _{r}}\left[ C_{n}-C_{p}%
\right]
\end{equation}
what after insertion of $\left( 76\right) $ into $\left( 80\right) $ gives 
\begin{equation}
\alpha =\frac{kT}{e}\frac{\left[ C_{n}-C_{p}\right] }{\left[ C_{n}+C_{p}%
\right] }
\end{equation}
Similarly as in the case of a semiconductor, one can write our equation in
the form 
\begin{equation}
\left[ \nabla ^{2}-\mu ^{2}\right] \phi ^{\prime }\left( x\right) =-4\pi 
\frac{\eta e}{4\pi \varepsilon _{0}\varepsilon _{r}}\delta \left( x\right)
\end{equation}
and arrive at the solution 
\begin{equation}
\phi \left( x\right) =\frac{\eta e}{4\pi \varepsilon _{0}\varepsilon _{r}x}%
\exp \left[ -\left( \frac{kT}{e}\frac{\left[ C_{n}-C_{p}\right] }{\left[
C_{n}+C_{p}\right] }\right) x\right] +\frac{kT}{e}\frac{\left[ C_{n}-C_{p}%
\right] }{\left[ C_{n}+C_{p}\right] }.
\end{equation}
It is worth to compare our approach with that described in Jackson's
monograph $\left[ 6\right] $ on p.342:

..''A nonrigorous derivation of the screening effect described above was
first given by Debye and Huckel in their theory of electrolytes. The basic
argument is as follows:

...''A nonrigorous derivation of the screening effect described above was
first given by Debye and Huckel in their theory of electrolites. The basic
argument is as follows. Suppose that we have a plasma with a distribution of
electrons in thermal equilibrium in an electrostatic potential $\Phi $. Then
they are distributed according to the Boltzmann factor $e^{-\frac{H}{kT}}$
where $H$ is the electric Hamiltonian. The spatial density of electrons is
therefore 
\begin{equation}
n\left( x\right) =n_{0}\exp \left[ -\frac{e\Phi }{kT}\right] .
\end{equation}
Now we imagine a test charge $Ze$ placed at the origin in this distribution
of electrons with its uniform background of positive ions (charge density $%
-en_{0}$). The resulting potential $\Phi $ will be determined by Poisson's
equation 
\begin{equation}
\nabla ^{2}\Phi =-4\pi Ze\delta \left( x\right) -4\pi en_{0}\left[ \exp
\left( -\frac{e\Phi }{kT}\right) -1\right] .
\end{equation}
If 
\begin{equation*}
\frac{e\Phi }{kT}
\end{equation*}
is assumed small, the equation can be linearized: 
\begin{equation}
\nabla ^{2}\Phi -k_{D}^{2}\Phi \backsimeq -4\pi Ze\delta \left( x\right)
\end{equation}
where 
\begin{equation}
k_{D}^{2}=\frac{4\pi n_{0}e^{2}}{kT}
\end{equation}
...Equation $\left( 87\right) $ has the spherically symmetric solution: 
\begin{equation}
\Phi \left( r\right) =Ze\frac{e^{-k_{D}r}}{r}
\end{equation}
showing that the electrons move in such a way to screen out the Coulomb
field of a test particle in a distance of the order of $k_{D}^{-1}$.''...

It can be seen that the approach described by Jackson contains less
parameters. The other difference is that in our approach one can state the
boundary conditions while Jackson considers an infinite system.

It is also worth to stress that in our approach it is possible to describe
the relaxation of the system towards the stationary state.

\section{\protect\bigskip Conclusions}

It would be interesting to solve our equations without linearisations and to
compare our results with the experimental results (for isothermal
conditions).

In principle, our approach can be applied also to the systems with the
temperature changing in space and time (compare $\left[ 2,3,4\right] $). We
hope to discuss that in future.

\qquad \bigskip

{\Large Acknowledgement} I am grateful to prof J.Lewandowski and
Dr.G.Radzikowska for their help.

\bigskip \qquad

{\Large Bibliography}

\bigskip

1. R.Peierls, Quantum theory of solids, Clarendon Press, Oxford, 1955.

2. S.Piekarski, On the modified Fick law and its potential applications,

J.Tech.Phys.,44,2,125-131,2003.

3. S.Piekarski, On diffusion and thermodiffusion in a gravity field,

J.Tech.Phys.,44,3,329-337,2003.

4.S.Piekarski, On the thermodiffusion of electrically charged matter.

In preparation.

5.V.L.Bonch - Brujewitsch,.S.G.Kalashnikov, Physics of semiconductors,

PWN, Warszawa, 1985 (in polish); translated from russian, Fizika

poluprowodnikov, Mockva, Hauka, 1977.

6.J.D.Jackson, Classical electrodynamics, John Wiley\&Sons, 1975.

7.W.Nowacki, Electromagnetic effects in deformable solid bodies,

PWN,Warszawa, 1983, in polish.

8.R.Wojnar, Nonlinear heat equation and thermodiffusion, 

Rep. on Math. Phys.,46,1/2, 2000.

9.R.Wojnar, On nonlinear heat equations and diffusion in porous media,

Rep. on Math. Phys.,44, 1/2, 1999.

\end{document}